\documentclass[arXiv]{actapoly}

\usepackage{xcolor}

\newcommand{\ii}{\mathrm{i}}	        % imaginary
\newcommand{\dd}{\mathrm{d}}		    % upshape d
\renewcommand{\vec}[1]{\boldsymbol{#1}}	% bold vectors
\renewcommand{\t}[1]{\text{#1}}	        % text

\begin{document}

\title[Numerical Calculation of the Complex Berry Phase]{Numerical Calculation of the Complex Berry Phase in Non-Hermitian Systems}

\correspondingauthor[M. Wagner]{Marcel Wagner}{}{marcel.wagner@itp1.uni-stuttgart.de}
\author[F. Dangel]{Felix Dangel}{}
\author[H. Cartarius]{Holger Cartarius}{}
\author[J. Main]{J\"org Main}{}
\author[G. Wunner]{G\"unter Wunner}{}
\shortauthors{M. Wagner et al.}

\institution{}{Institut f\"ur Theoretische Physik 1, Universit\"at Stuttgart,
  70550 Stuttgart, Germany}

\begin{abstract}
  We numerically investigate topological phases of periodic lattice systems in
  tight-binding description under the influence of dissipation. The effects
  of dissipation are effectively described by $\mathcal{PT}$-symmetric
  potentials. In this framework we develop a general numerical gauge smoothing
  procedure to calculate complex Berry phases from the biorthogonal basis of
  the system's non-Hermitian Hamiltonian. Further, we apply this method to a
  one-dimensional $\mathcal{PT}$-symmetric lattice system and verify our numerical 
  results by an analytical calculation.
\end{abstract}


\maketitle

\section{Introduction}
\label{sect:int}

Due to their robustness against local defects or disorder topologically
protected states as Majorana fermions \cite{Mourik2012a,Stanescu2013a,%
	Elliott2015a} are of high value for physical applications such as quantum
computation \cite{Stern2013a}. However, no physical system is completely
isolated and dissipation can have an important influence on the states
\cite{Carmele2015a}. Majorana fermions can even be created with the help of
dissipative effects \cite{Bardyn2013a,SanJose2016a}.

Of special importance in this context is the case of balanced gain and loss
as described by $\mathcal{PT}$-symmetric complex potentials \cite{Bender98},
which has attracted large interest in quantum mechanics \cite{Znojil2001a,
	Bender2002a,Znojil2011a,Jones2010a,Li2012a}. The stationary Schr\"odinger
equation was solved for scattering solutions \cite{Mehri-Dehnavi2010a} and
bound states \cite{Jakubsky2005}, and also questions concerning other
symmetries \cite{Levai2002a,Abt2015a} as well as the meaning of non-Hermitian
Hamiltonians have been discussed \cite{Mostafazadeh2006a,Jones2008a}. Their
influence on many-particle systems has been studied mainly in the context of
Bose-Einstein condensates \cite{Graefe08b,Graefe08a,Musslimani08b,Heiss13a,%
	Dast14a,Gutoehrlein15a} but also on lattice systems \cite{Hu2011a,Esaki2011a,%
	Yuce2015a,Yuce2015b,Yuce2016a,Klett2017a}. In the latter systems it was
shown that $\mathcal{PT}$-symmetric complex potentials may eliminate the
topologically protected states existing in the same system under complete
isolation \cite{Hu2011a,Esaki2011a,Yuce2015a,Yuce2015b,Yuce2016a,%
	Schomerus2013a,Zeuner2015a}. However, in some $\mathcal{PT}$-symmetric
potential landscapes they can survive \cite{Yuce2015a,Yuce2015b,Yuce2016a,%
	Klett2017a,Schomerus2013a,Wang15}, which has been confirmed in impressive
experiments \cite{Zeuner2015a,Weimann2016a}

In most theoretical studies the topologically protected states have been
identified via their property to close the energy gap of the band structure or
their localisation at edges or interfaces of the systems \cite{Hu2011a,%
  Schomerus2013a,Yuce2015a,Yuce2015b,Yuce2016a}. The identification and
calculation of topological invariants such as the Zak phase \cite{Zak1989}
known from Hermitian systems leads to new challenges in the case of non-Hermitian
operators \cite{Esaki2011a,Liang2013,Mandal2016a}. This is especially true if
the eigenstates are only available numerically. Indeed, in Refs. 
\cite{Esaki2011a,Liang2013,Mandal2016a} all calculations have been done
for eigenstates which are analytically available. However, for arbitrary
$\mathcal{PT}$-symmetric complex potentials analytical access to the eigenstates
is not available and a reliable numerical procedure is required. For the
calculation of the invariants an integration of phases over a
loop in parameter space is typically necessary. 

For example, in the case of the Zak phase,
which is applied to one-dimensional systems, this integral runs over the first
Brillouin zone. The integrand contains the eigenstates of the Hamiltonian and
their first derivatives. In a numerical calculation it is evaluated at
discrete points in momentum space and each state possesses an arbitrary
global phase spoiling the phase relation.

This is the point our study sets in. In this article we introduce
a robust method of calculating the complex Berry phase numerically. It is based
on a normalisation of the left- and right-hand eigen\-vectors with the
biorthogonal inner product \cite{Brody14}, which reduces to the c product
\cite{Moiseyev} in the case of complex symmetric Hamiltonians. To obtain an
unambiguous complex Berry phase we introduce a numerical gauge smoothing
procedure. It consists of two parts. First we have to remove the influence
of the arbitrary and unconnected global phases of the eigenstates, which is
unavoidably attached to them for each point in parameter space. This is achieved by relating the eigenvectors of consecutive steps in parameter space, and then normalising them. With this approach the eigenvectors are not yet single-valued, i.e. the vectors at the starting and end point of the loop possess different phases. These points have to be identified and will be refered as the basepoint of the loop later on. The phase difference between the different left and right states at the basepoint has to be corrected by ensuring the eigenvectors to be identical at the basepoint. 

The article is organised as follows. First, we introduce the complex
Berry phase in section \ref{sect:zak}. In section \ref{sect:ngs} we establish the
algorithm of the gauge smoothing procedure for non-Hermitian (and Hermitian)
Hamiltonians. An example is presented in section \ref{sect:Aplication}, where
we apply the previously developed method to a non-Hermitian extension of the
Su-Schrieffer-Heeger (SSH) model \cite{SSH79} to calculate its complex Zak phase.

\section{Complex Berry phase}
\label{sect:zak}
Topological phases of closed one-dimensional periodic lattice systems are
characterised by the Zak phase \cite{Zak1989}, which is the Berry phase
\cite{Berry1984} picked up by the eigenstate when it is transported once
along the Brillouin zone. In the presence of an antiunitary symmetry these
phases are quantised \cite{Hatsugai2006} and can be related to the winding
number of a vector $\vec{n}(k)$ determining the Bloch Hamiltonian
\begin{equation}
  \mathcal{H}(k)= \vec{n}(k) \cdot \vec{\sigma}\; ,
  \label{eq:bloch_winding}
\end{equation}
where $\vec{\sigma}$ denotes the vector of Pauli matrices and $k$ is the wave
number parametrising the Brillouin zone, which acts as parameter space. In
this case the Zak phase characterises the system's topological phase. 

This concept can be generalised to dissipative systems effectively described
by a $\mathcal{PT}$-symmetric non-Hermitian Hamiltonian $H(\vec{\alpha})$. The
complex Berry phase $\gamma_n$ of a biorthogonal pair of eigenvectors
$\langle \chi_n |$ and $|\phi_n \rangle$ of $H(\vec{\alpha})$, 
\begin{equation} \label{eq:Berry}
  \gamma_n= \ii \oint \limits_{\mathcal{C}} \, \langle \chi_n
  | \vec{\nabla}_{\hspace{-0.08cm}\vec{\alpha}} | \phi_n \rangle
  \cdot \dd\vec{\alpha} \; ,
\end{equation}
follows from the lowest order of the adiabatic approximation of the time
evolution of a state in parameter space \cite{Garrison1988}.
Here $\mathcal{C}$ is a  loop in parameter space and
\mbox{$\vec{\alpha}=(\alpha_1, ..., \alpha_{i}, ...)$} are its coordinates. We
consider $\mathcal{PT}$-symmetric non-Hermitian Hamiltonians of the form
\begin{equation}
H(\vec{\alpha})= H_{\t{h}}(\vec{\alpha}) + H_{\t{nh}}(\vec{\alpha}) \; ,
\end{equation}
where $H_{\t{h}}$ denotes its Hermitian and $H_{\t{nh}}$ represents its
$\mathcal{PT}$-symmetric non-Hermitian part. The non-Hermitian part is a
complex potential modelling the gain and loss of particles.

The complex Berry phase $\gamma_n$ arising from the periodic modulation of
states in the parameter space of a $\mathcal{PT}$-symmetric one-dimensional system cannot be related to a real winding number
calculated from eq.\ \eqref{eq:bloch_winding}. Hence, the calculation of the complex
Berry phase requires the determination of gauge-smoothed eigenvector pairs along
the loop $\mathcal{C}$ in parameter space allowing for the evaluation
of eq.\ \eqref{eq:Berry}.

Here, it is important to note that the argumentation of Hatsugai
\cite{Hatsugai2006} for the quantisation of Berry phases of Hermitian
Hamiltonians can be extended on complex Berry phases of non-Hermitian
$\mathcal{PT}$-symmetric Hamiltonians in the case of unbroken $\mathcal{PT}$
symmetry. One finds the real part of the complex Berry phase to take values
$0$ or $\pi$ modulo $2\pi$. Thus, a strict quantisation is still present and the $\mathcal{PT}$ symmetry protects the topological phases occurring in such systems.

\section{Numerical gauge smoothing}
\label{sect:ngs}

In this section we present a numerical procedure to determine the left and
right eigenvectors $\langle \chi_n |$ and $| \phi_n \rangle$ in an appropriate
smoothed gauge to compute complex Berry phases on a discretised loop
$\mathcal{C}=( \vec{\alpha}_1, ..., \vec{\alpha}_j, ..., \vec{\alpha}_M
=\vec{\alpha}_1 )$ in parameter space. This is necessary for the evaluation of integrals of the form in eq.\ \eqref{eq:Berry}. Typically the left and right eigenvectors have to be calculated independently,
and each of them has an arbitrary global phase. The biorthogonal normalisation
condition \cite{Brody14},
\begin{subequations}
\begin{align} \label{eq:normLeft}
	\langle \chi_n (\vec{\alpha}_j)| \rightarrow & \, \frac{\langle\chi_n (\vec{\alpha}_j)|}{ \sqrt{\langle \chi_n (\vec{\alpha}_{j})|\phi_n (\vec{\alpha}_{j}) \rangle} }\; ,\\
	\label{eq:normRight}
	|\phi_n (\vec{\alpha}_j) \rangle \rightarrow & \, \frac{|\phi_n (\vec{\alpha}_j) \rangle }{ \sqrt{\langle \chi_n (\vec{\alpha}_{j})|\phi_n (\vec{\alpha}_{j}) \rangle} } \; ,
\end{align}
\end{subequations}
chooses one arbitrary global phase for each $\vec{\alpha}_j$. This is sufficient if only products or matrix elements
of eigenstates belonging to the same point in parameter space are required.
However, for numerical derivatives used in eq.\ \eqref{eq:Berry} the remaining
global phases of successive steps in $\vec{\alpha}_j$ along the loop $\mathcal{C}$ can spoil the complex
Zak phase. A fixation of the phase between consecutive steps that does not
distort the desired result is required.

Starting point for the gauge smoothing procedure are the left and right handed versions of the
time-independent Schr\"odinger equation,
\begin{subequations}
\begin{align} \label{eq:lSG}
	\langle \chi_n (\vec{\alpha}) |\,H (\vec{\alpha}) = \, & E_n (\vec{\alpha}) \, \langle \chi_n (\vec{\alpha})|\; ,\\
	\label{eq:rSG}
	H(\vec{\alpha}) \,|\phi_n (\vec{\alpha}) \rangle = \, & E_n (\vec{\alpha})  \,|\phi_n (\vec{\alpha}) \rangle\; ,
\end{align}
\end{subequations}
defining a set of natural left and right basis states
$\langle \chi_n (\vec{\alpha})|$ and $|\phi_n (\vec{\alpha}) \rangle$.
These equations are solved for every point $\vec{\alpha}_j$ of the discretised loop $\mathcal{C}$ in parameter space providing the eigenvalues
$E_n (\vec{\alpha}_j)$ and the unnormalised states of a biorthogonal basis
$\{\langle \chi_n (\vec{\alpha}_j)|, |\phi_n (\vec{\alpha}_j) \rangle\}$ of the
Hamiltonian $H(\vec{\alpha}_j)$ at each point $\vec{\alpha}_j$. Here the basis
states are determined up to the aforementioned arbitrary phases.

To smooth the gauge within the loop in parameter space with basepoint $\vec{\alpha}_1$ one chooses an arbitrary global phase.
It is most convenient to do this for the basepoint. The corresponding eigenstates are normalised according to
the conditions \eqref{eq:normLeft} and \eqref{eq:normRight}. The following
two-stage procedure transfers the choice of the global phase at
$\vec{\alpha}_{1}$ onto the other basis states along the loop $\mathcal{C}$.

First one modifies the phases of the states $\langle \chi_n (\vec{\alpha}_j)|$
and $|\phi_n (\vec{\alpha}_j) \rangle$ iteratively by
\begin{subequations}
\begin{align}\label{eq:tracedLeft}
  \langle \chi_n (\vec{\alpha}_j)| \rightarrow & \, \langle \chi_n (\vec{\alpha}_j)| \,\, \t{e}^{-\ii \, \t{arg}(\langle \chi_n (\vec{\alpha}_j)|\phi_n (\vec{\alpha}_{j-1}) \rangle)}\; ,\\
  \label{eq:tracedRight}
  |\phi_n (\vec{\alpha}_j) \rangle \rightarrow & \, |\phi_n (\vec{\alpha}_j) \rangle \,\, \t{e}^{-\ii \, \t{arg}(\langle \chi_n (\vec{\alpha}_{j-1})|\phi_n (\vec{\alpha}_{j}) \rangle)}
\end{align}
\end{subequations}
followed by a normalisation of the states
according to eqs.\,\eqref{eq:normLeft} and \eqref{eq:normRight}.
Eqs.\ \eqref{eq:tracedLeft} and \eqref{eq:tracedRight} relate the vectors of step $j$ to those of step $j-1$ by ensuring
\begin{subequations}
  \begin{align} \label{eq:steprelation1}
    \t{Im}\big(\langle \chi_n (\vec{\alpha}_j)|\phi_n (\vec{\alpha}_{j-1}) \rangle \big)=\, \,  & 0 \; , \\
    \label{eq:steprelation2}
    \t{Im} \big(\langle \chi_n (\vec{\alpha}_{j-1})|\phi_n (\vec{\alpha}_{j}) \rangle \big)= \, \, & 0 \; ,
  \end{align}
\end{subequations}
which is a valid condition in the continuous limit. The normalisation conditions \eqref{eq:normLeft} and \eqref{eq:normRight} ensure that the basis states now fulfil
\begin{equation}
\langle \chi_m (\vec{\alpha}_j)|\phi_n (\vec{\alpha}_j) \rangle = \delta_{mn}
\end{equation}
for $j \in \{1, ..., M\}$ and for all $n$ and $m$.

As a result of the first step the arbitrary
global phases have been removed. Only one arbitrary phase is left, which has
no influence, since it is identical for all right eigenvectors and its complex
conjugate for all left eigenvectors. However, the biorthogonal basis following
from the procedure so far is not single-valued in the parameter space. In
particular, the vectors at the starting and end point of the loop are not identical.
For the calculation of a Berry phase a continuous single-valued phase function
is essential \cite{Berry1984}, and thus has to be established.

To this end in the second step one adjusts the basis states such that they
are the same at the starting and the end point of the loop. This can be
achieved by compensating the phase difference between the states at the
basepoint, $\langle \chi_n (\vec{\alpha}_1)|$ and $\langle \chi_n
(\vec{\alpha}_M)|$, respectively, as well as $|\phi_n (\vec{\alpha}_1) \rangle$
and $|\phi_n (\vec{\alpha}_M) \rangle$. This remains true for single vector components. Therefore we calculate the phase difference of the first
non-vanishing component $p$ of the left basis states $\langle \chi_n
(\vec{\alpha}_1)|$ and $\langle \chi_n (\vec{\alpha}_M)|$,
\begin{equation} \label{eq:Dphi}
  \Delta \varphi_n = \varphi_{n,M} -\varphi_{n,1}  + 2 \pi X_n\; ,
\end{equation}
where $\varphi_{n,j}=\t{arg}\big(\langle \chi_n (\vec{\alpha}_j)|_p\big)$ is
the argument of component $p$ of the left eigenvector at the point
$\vec{\alpha}_j$ in parameter space and $X$ denotes the sum
of directed crossings of the phase $\varphi_{n,j}$ over the borders of the standard
interval $[-\pi,\pi)$. Starting with $X=0$ we increase $X$ by one for every
jump of $\varphi_{n,j}$ from $-\pi$ to $\pi$ and
subtract $1$ for the opposite direction.

The states of the biorthogonal basis can then finally be gauge-smoothed by multiplying the states at $\vec{\alpha}_j$ by a phase factor according to
\begin{subequations}
  \begin{align} \label{eq:smoothedLeft}
    \langle \chi_n (\vec{\alpha}_j)| \rightarrow & \, \langle \chi_n (\vec{\alpha}_j)| \,\, \t{e}^{-\ii \, f_{\Delta \varphi_n}((j-1)/(M-1))}\; ,\\
    \label{eq:smoothedRight}
    |\phi_n (\vec{\alpha}_j) \rangle \rightarrow & \, |\phi_n (\vec{\alpha}_j) \rangle \,\, \t{e}^{~\ii \,f_{\Delta \varphi_n}((j-1)/(M-1))}
  \end{align}
\end{subequations}
for $j \in \{1, ..., M\}$, where $f_{\Delta \varphi_n}(x)$ is any ``smooth''
real valued continuous function 
\begin{subequations}
\begin{align} \label{eq:gaugeFunction}
  f_{\Delta \varphi_n}: ~~ [0, 1] ~ & \to ~~~ \mathbb{R}\\
  \begin{split}
  \label{eq:cond}
  \t{fulfilling:}  \hspace*{0.627cm}  0 ~~~ & \mapsto ~ f_{\Delta \varphi_n}(0)= 0\\
  1 ~~~ & \mapsto ~ f_{\Delta \varphi_n}(1)=  \, \Delta \varphi_n  \pm  2z\pi 
  \end{split}
\end{align}
\end{subequations}
with $z \in \mathbb{Z}$. Its explicit form is not critical since it only has to correct the total
phase change over the whole range of the loop. However, a linear progression
of the phase correction from step to step turns out to be a good choice.

It should be mentioned that in case of a degeneracy of the eigenvalue at
$\vec{\alpha}_j$ the solution of eqs.\ \eqref{eq:lSG} and \eqref{eq:rSG} yields
an arbitrary linear combination of eigenvectors of the degenerate eigenspace.
To find the correct eigen\-vectors $\langle \chi_n (\vec{\alpha}_j)|$ and
$| \phi_n (\vec{\alpha}_j) \rangle$ one can apply a biorthogonal Gram-Schmidt
algorithm \cite{Parlett1985}. If $\vec{\alpha}_{j-1}$ is a point
neighbouring the degeneracy one tries to find a linear combination of the
vectors of the left degenerate eigenspace fulfilling
\begin{equation}
  \langle \chi_m (\vec{\alpha}_j)|\phi_n (\vec{\alpha}_{j-1}) \rangle \approx \delta_{mn} 
\end{equation}
and then chooses the right eigenvectors such that
\begin{equation}
  \langle \chi_m (\vec{\alpha}_j)|\phi_n (\vec{\alpha}_{j}) \rangle = \delta_{mn} \; .
\end{equation}
Alternatively one can treat the real and imaginary parts of the degenerate
eigenvector components as ``smooth'' functions. Then the eigenvector components
at degeneracy points can be predicted by fitting a spline to the vector
components at neighbouring points  $\vec{\alpha}_l$. An approximation to the correct eigenvectors of the
degenerate eigenspace can be determined by a linear combination of the obtained
vectors of the degenerate eigenspace such that they fit best to the prediction.
Hermitian Hamiltonians can be treated as a special case, in which the left
eigenvector fulfils $\langle \chi_n | = (| \phi_n \rangle^{\t{T}})^{\ast}$.

\section{Application to a one-dimensional lattice system}
\label{sect:Aplication}

In this section we apply the gauge smoothing procedure developed in section \ref{sect:ngs} to a $\mathcal{PT}$-symmetric one-dimensional lattice system to calculate the complex Zak phase
\begin{equation} \label{eq:Zak}
\gamma_n= \oint \limits_{\mathcal{BZ}} \langle \chi_n | \partial_k | \phi_n \rangle \, \dd k \; ,
\end{equation}
where the parameter space is given by the discretised Brillouin zone $\mathcal{BZ}$ and $k$ is the wave number.

\begin{figure}
	\centering                            
	\includegraphics[width=\linewidth]{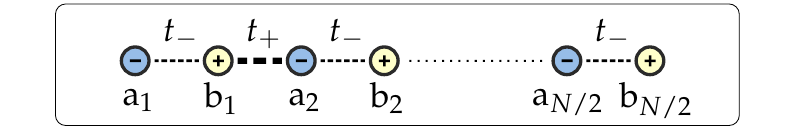} % <-- use this for your graphics
	\caption{(Colour online) Sketch of the SSH model with $N$ lattice sites subject to the alternating imaginary potential $U$. The minus (plus) sign marks a negative (positive) imaginary potential corresponding to particle sinks (sources).}
	\label{fig:SSHU2}         
\end{figure}

As an example we consider the SSH model \cite{SSH79} with $N$ lattice sites, tunnelling amplitudes $t_+$ and $t_-$, and creation (annihilation) operators of spinless fermions $c_n^{\dagger}$ $(c_{n}^{{\color{white}{\dagger}}})$ at site $n$,
\begin{align}
	\begin{split}
	H_{\t{SSH}}=&\sum \limits_{n=1}^{N/2} \,  t_- \left(c_{\t{a}_n}^{\dagger}\, c_{\t{b}_n}^{{\color{white}{\dagger}}} + \t{h.c.} \right) \\
	&+ \sum \limits_{n=1}^{N/2-1} \, t_+ \left(c_{\t{b}_n}^{\dagger}\, c_{\t{a}_{n+1}}^{{\color{white}{\dagger}}} + \t{\t{h.c.}} \right) \; .
	\end{split}
\end{align}
We apply an alternating non-Hermitian $\mathcal{PT}$-sym\-metric potential of the form
\begin{equation}
U=\ii \, \frac{\varGamma}{2} \sum \limits_{n=1}^{N/2} \left( c_{\t{b}_n}^{\dagger}c_{\t{b}_n}^{{\color{white}{\dagger}}} -\, c_{\t{a}_n}^{\dagger}c_{\t{a}_n}^{{\color{white}{\dagger}}}\right)~,
\end{equation}
where $\varGamma$ denotes the parameter of gain and loss. The $\mathcal{PT}$-symmetric Hamiltonian describing this model (sketched in Figure \ref{fig:SSHU2}) is given by 
\begin{equation} \label{eq:SSHU2Ham}
	H=H_{\t{SSH}}+U \; .
\end{equation}

To evaluate eq.\ \eqref{eq:Zak} we need to represent this Hamiltonian in the reciprocal space, where the Brillouin zone acts as parameter space. This is done by rewriting the Hamiltonian with creation and annihilation operators in the reciprocal space in the limit $N \to \infty$,
\begin{align} \label{eq:Bloch}
	H=\sum \limits_{k=-\pi}^{\pi} & \begin{pmatrix}
	c_{\t{a},k}^{\dagger}\, , \, c_{\t{b},k}^{\dagger}
	\end{pmatrix} \\
	\nonumber
	& 
	{\begin{pmatrix}
			-\ii \, \varGamma/2 & t_-+t_+ \t{e}^{\,\ii k} \\
			t_-+t_+ \t{e}^{-\ii k} & \ii \, \varGamma/2
			\end{pmatrix}} 
	\begin{pmatrix}
	\vspace{0.25cm}c_{\t{a},k}^{{\color{white}{\dagger}}}\\ 
	\vspace{0.1cm}c_{\t{b},k}^{{\color{white}{\dagger}}}
	\end{pmatrix}~,
\end{align}
where the sum runs over discrete values of $k$ in steps of $k=2\pi/N$ and the annihilation operator of an electron with wave number $k$ is given by
\begin{equation}
	c_n^{{\color{white}{\dagger}}}=\frac{1}{\sqrt{N}} \sum \limits_{k} c_{k}^{{\color{white}{\dagger}}} \t{e}^{-\ii \, k r_n} 
\end{equation}
with $r_n=an$ and the lattice spacing $a$. The matrix occurring in eq.\ \eqref{eq:Bloch} is the Bloch Hamiltonian $\mathcal{H}(k)$ of the system, which can be decomposed into the Pauli matrices,
\begin{align}
%\begin{split}
	\mathcal{H}(k) & =  \big(t_- +t_+ \t{cos}(k)\big) \sigma_1 -t_+ \t{sin}(k) \sigma_2 -\ii \Gamma/2 \sigma_3 \nonumber \\
	& = \vec{\mathfrak{n}}(k) \cdot \vec{\sigma}\;
%\end{split}\hspace*{-0.5cm}
\end{align}
with a coefficient vector $\vec{\mathfrak{n}}$ and the Pauli vector $\vec{\sigma}$. From this form the energy eigenvalues can be obtained explicitly,
\begin{equation}
	E_{\pm}(k) =\pm |\vec{\mathfrak{n}}(k)|\; .
\end{equation}

In the limit $\Gamma \rightarrow 0$ the Hamiltonian from eq.\ \eqref{eq:Bloch} reproduces the Hermitian SSH model, which possesses time-reversal, reflection, particle-hole, and a chiral symmetry (represented by $\sigma_3$). The introduction of a $\mathcal{PT}$-symmetric non-Hermitian on-site potential $\Gamma$ breaks these symmetries. The non-Hermitian Bloch Hamiltonian is invariant under the combined action of the parity and the time inversion operator. Further the particle-hole symmetry is broken because the sources (sinks) of an electron correspond to sinks (sources) of holes. The non-Hermitian system is therefore symmetric under the action of the combination of the parity and the charge conjugation operator. Further it has no chiral symmetry $\Lambda = \mathfrak{a}_0 \sigma_0 + \vec{\mathfrak{a}} \cdot \vec{\sigma}$ because a chiral symmetry would fulfil
\begin{align}
			\{  \Lambda, \mathcal{H}\} & = \sum_{i=0}^{3} \sum_{j=1}^{3} \mathfrak{a}_i \mathfrak{n}_j \{ \sigma_i , \sigma_j \} \\ 
			\nonumber & = 2 \big( \mathfrak{a}_1 \mathfrak{n}_1 + \mathfrak{a}_2 \mathfrak{n}_2 +\mathfrak{a}_3 \mathfrak{n}_3 \big) \sigma_0 + 2 \mathfrak{a}_0 \sum_{j=1}^{3} \mathfrak{n}_j \sigma_j \\
			\nonumber & \overset{!}{=} 0 
\end{align}
with a coefficient vector $\vec{\mathfrak{a}}$ which is independent of the value of $k$ , the $2 \times 2$ identity matrix $\sigma_0$, and the anti-commutation relations $\{ \sigma_i , \sigma_j \} = 2 \delta_{ij} \sigma_0$ of the Pauli matrices. Therefore one finds $\mathfrak{a}_0 = 0$, and thus
\begin{equation}
	\mathfrak{a}_1 \mathfrak{n}(k)_1 + \mathfrak{a}_2 \mathfrak{n}(k)_2 +\mathfrak{a}_3 \mathfrak{n}_3 \overset{!}{=} 0\; ,
\end{equation}
which cannot be satisfied for a constant vector $\vec{\mathfrak{a}}$ because the vector $\vec{\mathfrak{n}}(k)$ rotates on a cylindrical surface as $k$ runs through the Brillouin zone. Hence the non-Hermitian Hamiltonian in eq.\ \eqref{eq:Bloch} does not possess a chiral symmetry. However, this does not mean there is no quantised real part of the Zak phase since its quantisation is ensured by the argument of Hatsugai \cite{Hatsugai2006} in the $\mathcal{PT}$-unbroken parameter regime as mentioned above.  
At the critical point $\Gamma=0$ the system reproduces the Hermitian SSH model, which possesses the previously mentioned symmetries. For $\Gamma<0$ the particle sinks and sources are interchanged leading to a spatially reflected system with the same general properties as the system with $\Gamma>0$.

From the Bloch Hamiltonian (cf.\ eq.\ \eqref{eq:Bloch}) the complex Zak phase can be calculated following the steps explained in section \ref{sect:ngs}. We choose (c.f. eq.\ \eqref{eq:gaugeFunction})
\begin{align}
\begin{split}
	f_{\Delta \varphi_n}(x) \,=~ & \Delta \varphi_n \, x -2\pi \\
	\t{with}~~~x~=~ & \frac{k+\pi/a}{2\pi/a} \; ,
\end{split}
\end{align} 
which is the most simple function fulfilling the conditions \eqref{eq:cond}.

\begin{figure}
	\centering
	\includegraphics[width=\linewidth]{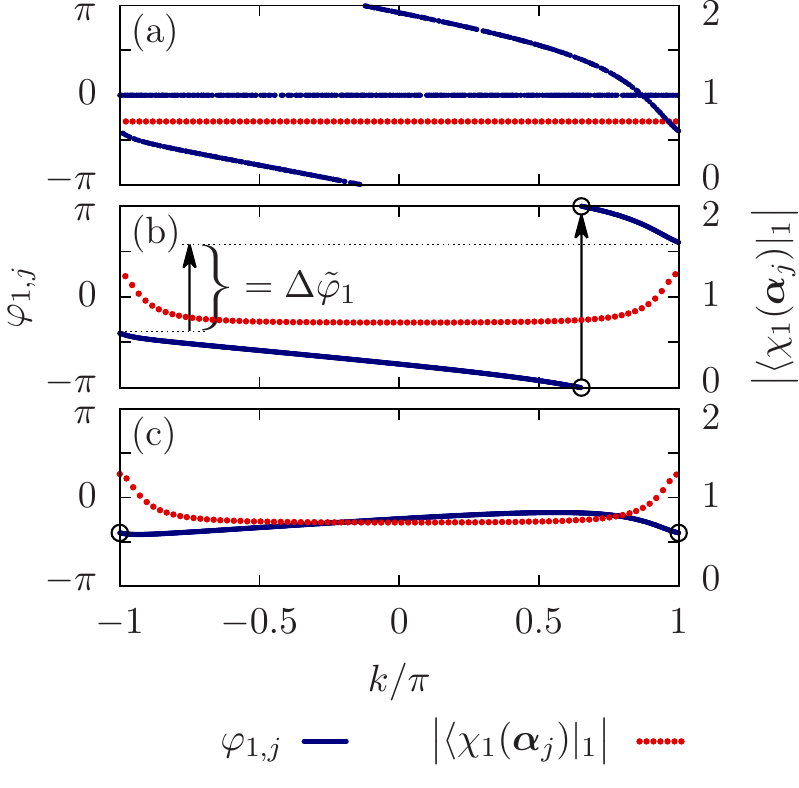} % <-- use this for your graphics
		\caption{(Colour online) First component of the left handed eigenvector $\langle \chi_1 (\vec{\alpha}_j)|_1$ in dependence of the wave number $k$ with $t=1$, $\Delta=1/2$, $\Gamma=1$ and $\theta \approx 0.3 \pi$\,: (a) Before the steps described in eqs.\ \eqref{eq:tracedLeft} and \eqref{eq:tracedRight} one can identify two different phase branches (blue line) and a constant modulus (filled red dots). (b) After the steps characterised by eqs.\ \eqref{eq:normLeft} and \eqref{eq:normRight} (here $\Delta \tilde{\varphi}_1=\varphi_{M,1} -\varphi_{1,1}$ and $X=-1$ c.f. eq.\ \eqref{eq:Dphi}) the phase is smooth within the Brillouin zone but discontinuous at its boundaries, and the modulus varies with $k$. (c) After the gauge smoothing process the phase difference $\Delta \tilde{\varphi}_1$ is compensated and the phase is continuous and smooth in the whole Brillouin zone.}
	\label{fig:CalcIll}
\end{figure}

Figure \ref{fig:CalcIll} illustrates the gauge smoothing process using the first
non-vanishing component $p=1$ of the left basis states as an example. The component $\langle \chi_1 (\vec{\alpha}_j)|_1$ of the unnormalised left eigenvector is shown in Figure\ \ref{fig:CalcIll}~(a). One can see two different phase branches as a result of the numerical diagonalisation and a constant modulus. After the gauge smoothing and normalisation according to the first step described in section \ref{sect:ngs} as shown in Figure\ \ref{fig:CalcIll}~(b) the modulus varies with the wave number $k$ and there is only one phase branch left, but the basis is not yet single-valued as there is still a phase difference at the boundaries of the Brillouin zone. In this example the factor $X=-1$ has to be used as $\varphi_{1,j}$ jumps from $-\pi$ to $\pi$. After the second step the component $\langle \chi_1 (\vec{\alpha}_j)|_1$ of the final left eigenvector is the same at the Brillouin zone boundaries in Figure\ \ref{fig:CalcIll}~(c).

\begin{figure}
	\centering
	\includegraphics[width=\linewidth]{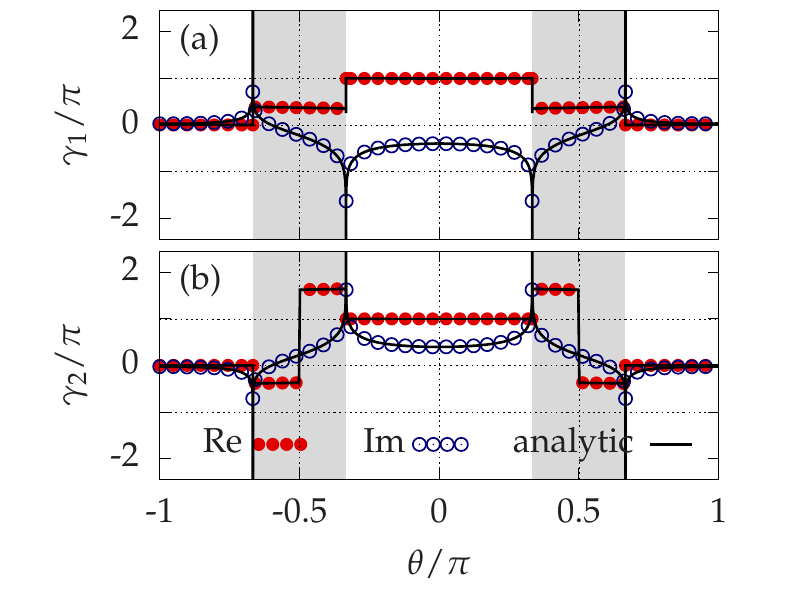} % <-- use this for your graphics
	\caption{(Colour online)  Numerical results of the real part (filled red dots) and the imaginary part (open blue circles) of the complex Zak phases $\gamma_1$ (a) and $\gamma_2$ (b) following from the Hamiltonian given in eq.\ \eqref{eq:SSHU2Ham} in dependence of the control parameter $ \theta$ with $t=1$, $\Delta = 1/2$ and $\Gamma=1$ (all values in a.u.) compared to the analytical result (real and imaginary part are represented by a solid black line coinciding with the numerical results) following from eq.\ \eqref{eq:analytic}. The grey shaded area marks the $\mathcal{PT}$-broken phase of the Bloch Hamiltonian.}
	\label{fig:ZakSSHU2}
\end{figure}

The two complex Zak phases $\gamma_1$ and $\gamma_2$ following from the eigenvector pairs of $\mathcal{H}(k)$ are shown in Figures \ref{fig:ZakSSHU2} (a) and (b), where the control parameter $\theta$ is used to describe the difference between the two tunnelling amplitudes $t_+$ and $t_-$,
\begin{equation}
	t_{\pm}=t \, \big( 1 \pm \Delta \t{cos}(\theta) \big) 
\end{equation}
with the mean value of the tunnelling amplitude $t$ and the dimerization strength $\Delta$. 

To verify our results we compare them with the analytical ones derived in \cite{Liang2013},
\begin{equation}\label{eq:analytic}
	\gamma_{1/2} = \pi \, \Theta (q-1) \pm \ii \frac{\eta}{2} \sqrt{\frac{\nu}{q}} \left( K(\nu) + \frac{q-1}{q+1} \Pi(\mu, \nu) \right) \; ,
\end{equation}
where $q=t_+/t_-$ is the ratio of the tunnelling amplitudes, $\eta=\varGamma / (2t_-)$ and $\nu = 4q/((q+1)^2-\eta^2)$ and $\mu=4q/(q+1)^2$. $K(\nu)$ and $\Pi(\mu, \nu)$ are elliptic integrals of first and third kind,
\begin{align}
	K(\nu)= & \int \limits_{0}^{\pi/2}\frac{\dd k}{\sqrt{1-\nu \, \t{sin}^2(k)}} \; , \\
	\Pi(\mu, \nu)= &\int \limits_{0}^{\pi/2} \frac{\dd k}{1-\mu \, \t{sin}^2(k) \, \sqrt{1-\nu \, \t{sin}^2(k)}} \; . 
\end{align}

The numerical calculations perfectly reproduce the analytical results. The grey shaded areas in Figure \ref{fig:ZakSSHU2} mark the values of $\theta$ for which the system is in a $\mathcal{PT}$-broken phase, for all other values of $\theta$ the system is in a $\mathcal{PT}$-unbroken phase. In the $\mathcal{PT}$-symmetric regime the real part of the complex Zak phase is either $0$ or $\pi$ modulo $2\pi$ and can be used to characterise the topological phase of the system.

\section{Conclusion}

We developed a numerical method to determine a gauge-smoothed biorthogonal basis of eigenstates of a $\mathcal{PT}$-symmetric non-Hermitian Hamiltonian required for complex Zak phases. It is also applicable to Hermitian systems. In the course of this we removed the arbitrary and unconnected global phases of the biorthogonal eigenstates of the $\mathcal{PT}$-symmetric Hamiltonian at each point in parameter space and made the basis single-valued. This allows for the calculation of the complex Berry phase by explicitly evaluating eq.\ \eqref{eq:Berry} even in complicated lattice systems for which no analytical access to the eigenstates is approachable. We demonstrated the action of the gauge smoothing method by applying the developed algorithm to a $\mathcal{PT}$-symmetric extension of the SSH model. An excellent agreement of the numerical and analytical results was found. In future work this provides the basis for the identification of the Zak phase as topological invariant in many-body systems with arbitrary $\mathcal{PT}$-symmetric complex potentials.

%\bibliographystyle{actapoly}
%\bibliography{biblio}

\end{document}